\documentclass{cjaa}                   

\usepackage{graphicx}                   
\input{epsf.sty}                        
\input{psfig.sty}                       

\begin{document}

   \title{The Black Hole Mass and Magnetic Field Correlation in Active
   Galactic Nuclei
}

   \volnopage{Vol.0 (200x) No.0, 000--000}      
   \setcounter{page}{1}          

   \author{W.M.Zhang
      \inst{1}\mailto{}
   \and Y.Lu
      \inst{2}
   \and S.N.Zhang
      \inst{1}
      }
   \offprints{A.-Y. Zhou}                   

   \institute{Physics Department and Center for Astrophysics, Tsinghua University,
             Beijing 100084, China\\
             \email{Zhangweiming@mails.tsinghua.edu.cn}
        \and
             National Astronomical Observatories, Chinese Academy of Sciences,
             Beijing 100012, China\\
          }

   \date{Received~~2001 month day; accepted~~2001~~month day}

   \abstract{  The observed $\nu$\emph{L}$_{\nu}$(5100\AA)-\emph{M}$_{BH}$ correlation is used to probe  the magnetic field of black holes harbored in active
galactic nuclei (AGNs). The model is based on the assumption that
the disk is heated by energy injection due to the magnetic
coupling (MC) process and the gravitational dissipation due to
accretion. The MC process can transfer energy and angular momentum
from a rotating Kerr black hole to its surrounding disk. The
relation of $\nu$\emph{L}$_{\nu}$(5100\AA)-\emph{M}$_{BH}$ as
functions of the spin ($a_*$) and magnetic field ($B_{BH}$) of the
black hole (BH) is modelled. The model predicts that $\nu
L_\nu(5100\AA)$ emitted from the disk is sensitive to the strength
of the poloidal component of the magnetic field on the BH horizon.
Based on the observations of $\nu L_\nu(5100\AA)$ for 143 AGN
sources, we obtain the correlation between ${M}_{BH}$ and
$B_{BH}$. And we compared our result with the approximate result
between ${M}_{BH}$ and $B_{BH}$ derived from the condition in the
standard accretion disc theory. 
   }

   \authorrunning{W. M. Zhang, Y. Lu \& S. N. Zhang }            
   \titlerunning{The Black Hole Mass and its Magnetic Field in Active Galactic Nuclei}  

   \maketitle

%
%
\section{Introduction}           
\label{sect:intro}

Recently, as a possible mechanism for transferring energy and
angular momentum from a fast-rotating BH to its surrounding disk,
the magnetic coupling (MC) process through the closed field lines
connecting the Kerr BH with its exterior disk, has been widely
studied (Li.~\cite{li02a}; Wang et
al.~\cite{wang02},~\cite{wang03}), which can be regarded as one of
the variants of the Blandford-Znajek (BZ) process proposed two
decades ago (Blandford \& Znajek.~\cite{blan77}). This mechanism
has been applied to account for the recent XMM-Newton observation
of the nearby bright Seyfert 1 galaxy MCG-6-30-15
(Li.~\cite{li02b}) which has a very steep emissivity. It is
assumed that the disk is stable, perfectly conducting, thin and
Keplerian. The magnetic field is assumed to be constant on the
black hole horizon and to vary as a power law with the radial
coordinate of the disk.

The model involving the magnetic field configurations with both
poloidal and toroidal components still remains an open question.
Since the magnetic field $B_{BH}$ on the BH horizon is brought and
held by its surrounding magnetized disk, there must exist some
relations between the magnetic field and the accretion rate.
However, studying the structure of $B_{BH}$ from its relation with
the accretion rate is difficult because estimation of the
accretion rate from observations is not easy (Wang 2003).
Motivated by recent studies on the BH mass in AGNs and inactive
galaxies, the relation of
$\nu$\emph{L$_{\nu}(5100\AA)$}-\emph{M$_{BH}$} is used to
investigate the magnetic field of a BH.


\section{Model predictions of the $\nu$\emph{L$_{\nu}(5100\AA)$}-\emph{M$_{BH}$} correlation}
\label{sect:rel}

\begin{figure}[h]
   \vspace{2mm}
   \begin{center}
   \hspace{3mm}\psfig{figure=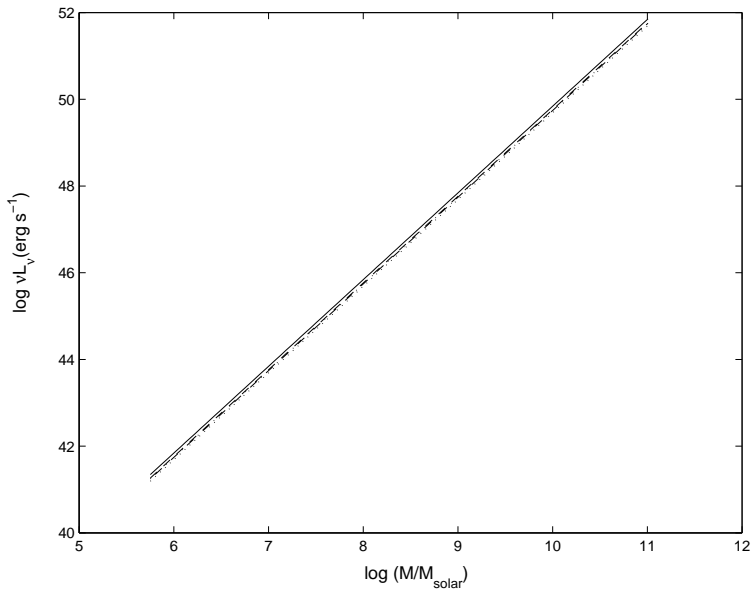,width=100mm,angle=0.0}
   \caption{{\small $\nu$\emph{L$_{\nu}$} versus \emph{M$_{BH}$}
  relationship for different values of black hole spin
  parameter $a_{*}$
with $n=3$ and $B=10^4$ G. The dotted, dashed-dotted, dashed and
solid lines correspond to \emph{a$_{*}$}=0.2, 0.5, 0.8 and 0.995,
respectively.} }
  \end{center}
\end{figure}
\begin{figure}[h]
   \vspace{2mm}
   \begin{center}
   \hspace{3mm}\psfig{figure=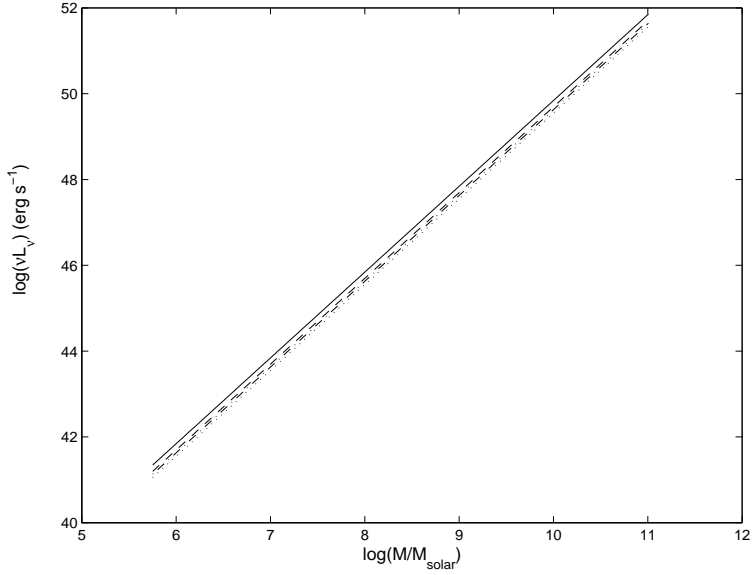,width=100mm,angle=0.0}
  \caption{{\small $\nu$\emph{L$_{\nu}$} versus \emph{M$_{BH}$}
  relationship for different values of the power-law index $n$ of magnetic field
  distribution (Eq. (9))
with $a_*=0.5$, and $ B=10^4\,G$. The dotted, dashed-dotted,
dashed and solid lines correspond to $n=1.5$, $3.0$, $5.0$ and
$7.0$, respectively.}}
   \end{center}
\end{figure}
\begin{figure}[h]
   \vspace{2mm}
   \begin{center}
   \hspace{3mm}\psfig{figure=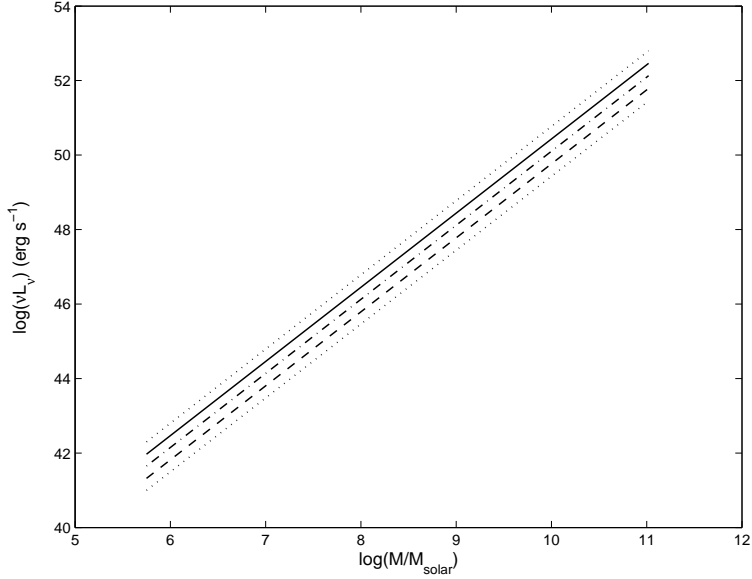,width=100mm,angle=0.0}
   \caption{ $\nu$\emph{L$_{\nu}$} versus \emph{M$_{BH}$}
  relationship for different values of black hole magnetic field
  $B_{BH}$
in the case of $a_*=0.5$ and $n=3.0$. The dotted, dashed,
dashed-dotted, solid and upper dotted lines correspond to
$B_{BH}=3\times$10$^{3}$ G, 1$\times$10$^{4}$ G, 3$\times$10$^{4}$
G, 1$\times$10$^{5}$ G and 3$\times$10$^{5}$ G, respectively
 }
   \label{Fig:lightcurve-ADAri}
   \end{center}
\end{figure}

\begin{figure}[h]
   \vspace{2mm}
   \begin{center}
   \hspace{3mm}\psfig{figure=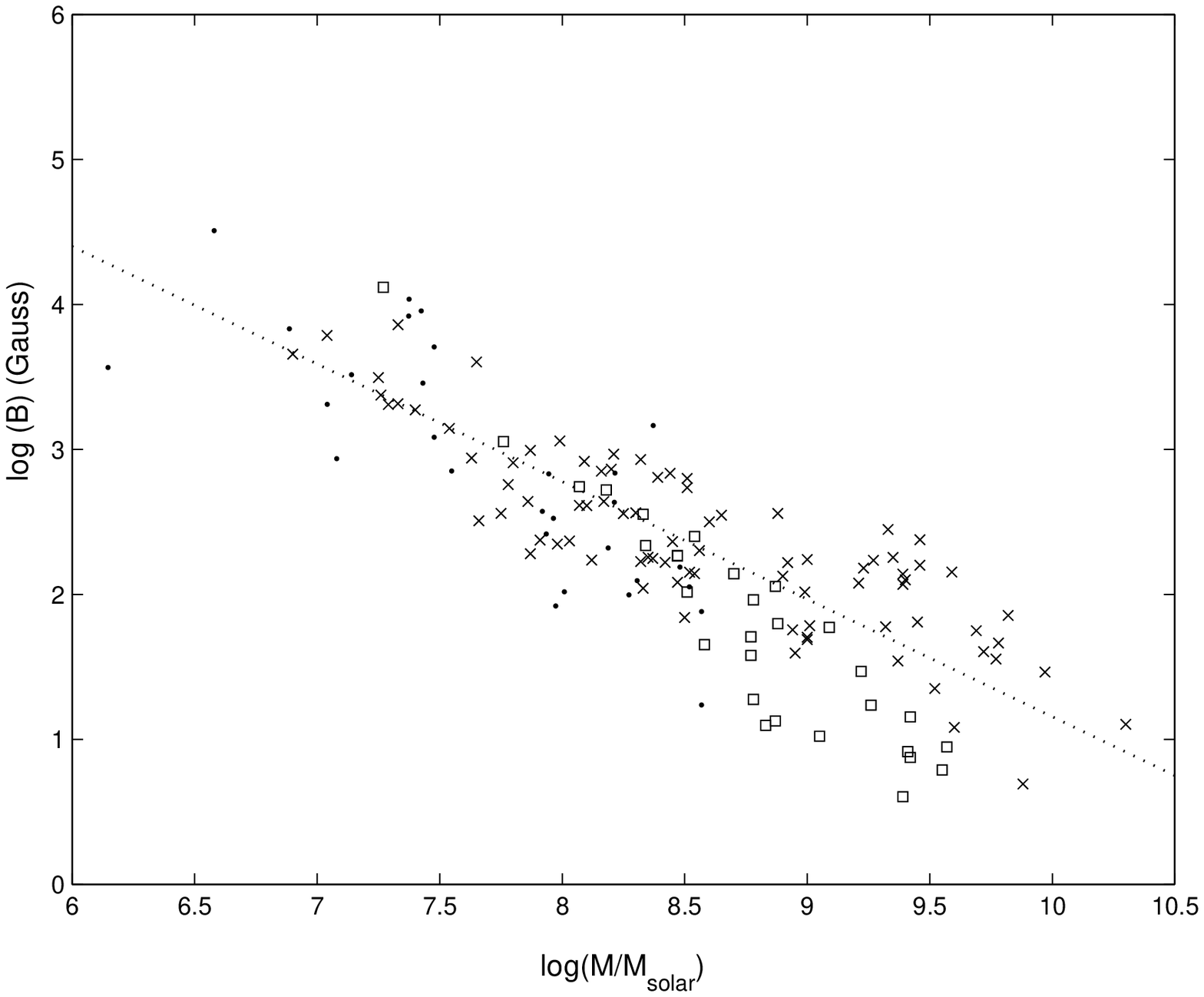,width=100mm,angle=0.0}
   \caption{ The derived \emph{B} versus \emph{M$_{BH}$} correlation. The sources are from Kaspi et
al.~\cite{kasp00}, McLure et al.~\cite{mclu01} and Shields et
al.~\cite{shie03}, corresponding to dots, squares and crosses,
respectively. The dashed-line is the best-fit result from the data
points, as shown in Eq. (13).}
   \label{Fig:lightcurve-ADAri}
   \end{center}
\end{figure}

It is known that the energy output from an accretion disk strongly
depends on its structure, particularly, the amounts of thermal
excess in the spectrum of the escaped radiation strongly depends
on the disk temperature. Assuming that the disk is heated by
considering the MC effects and the gravitational energy
dissipation, the disk temperature can be estimated. In this case,
the temperature is approximated by balancing absorption with
thermal emission
\begin{equation}
  \sigma T^{4}=F_{MC}+F_{DA}\,,
\label{eq:LebsequeI}
\end{equation}
where \emph{F$_{MC}$} and \emph{F$_{DA}$} are the emission fluxes
due to the MC effects (Wang et al.~\cite{wang02},~\cite{wang03})
and the disk accretion in the situation of no-torque boundary
condition (Page \& Thorne.~\cite{page74}), respectively. We assume
that thermal emission dominate the emission. Though there may
exist other radiation mechanisms, however the total amount of
$F_{MC}$ and its proportion in the total emission flux will not be
changed by different radiation mechanisms. The \emph{F$_{MC}$} and
\emph{F$_{DA}$} are formulated by
\begin{eqnarray}
 F_{DA}(x)&=& \frac{3\dot{M}c^{6}}{8\pi
  M^{2}G^{2}}\frac{1}{x^{4}(x^{3}-3x+2a_{*})}
  \times[x-x_{ms}-\frac{3}{2}a_{*}ln(\frac{x}{x_{ms}})-\frac{3(x_{1}-a_{*})^{2}}{x_{1}(x_{1}-x_{2})(x_{1}-x_{3})}\ln(\frac{x-x_{1}}{x_{ms}-x_{1}})\nonumber\\
 &-&\frac{3(x_{2}-a_{*})^{2}}{x_{2}(x_{2}-x_{1})(x_{2}-x_{3})}\ln(\frac{x-x_{2}}{x_{ms}-x_{2}})
  -\frac{3(x_{3}-a_{*})^{2}}{x_{3}(x_{3}-x_{1})(x_{3}-x_{2})}\ln(\frac{x-x_{3}}{x_{ms}-x_{3}})]\,\,,\\
  F_{MC}(x)&=&\frac{3}{2}\frac{Gx_{ms}^{4}}{Mc^{2}x(x^{3}+a_{*})^{2}}[E^{+}-\Omega_{D}L^{+}]^{-2}
  \times\int_{1}^{\xi_{out}}(E^{+}-\Omega_{D}L^{+})H\xi d\xi\,\,.
\label{eq:LebsequeI}
\end{eqnarray}
where $\xi$$_{out}$ is the outer boundary of the disk defined in
the later formula of Eq.(7).
\begin{eqnarray}
&&  A_{1}= 1+(1-a_{*}^{2})^{1/3}[(1+a_{*})^{1/3}+(1-a_{*})^{1/3}]
\,,\,\,\, A_{2}=(3a_{*}^{2}+A_{1}^{2})^{1/2}\,\,,\nonumber\\
&& x_{ms}= \{3+A_{2}+[(3-A_{1})(3+A_{1}+2A_{2})]^{1/2}\} \,,\,\,
  x_{1}=2\cos[\frac{1}{3}\arccos(a_{*})-\pi/3]\,\,,\nonumber\\
&&  x_{2}= 2\cos[\frac{1}{3}\arccos(a_{*})+\pi/3] \,,\,\,
  x_{3}=-2\cos[\frac{1}{3}\arccos(a_{*})]\,,\,\,\,
  \Omega_{D}=\frac{1}{M(x^{3}+a_{*})}\,\,,\\
&& E^{+} =
\frac{(1-2x^{-2}+a_{*}x^{-3})}{(1-3x^{-2}+2a_{*}x^{-3})^{1/2}}\,,\,\,\,
  L^{+}=\frac{Mx(1-2a_{*}x^{-3}+a_{*}^{2}x^{-4})}{(1-3x^{-2}+2a_{*}x^{-3})^{1/2}}\,\,.
\end{eqnarray}

The definitions of the basic parameters of a BH are:
\begin{eqnarray}
&& a=J/Mc\,,\,\,\, r_{g}=GM/c^{2}\,,\,\,\, a_{*}=a/r_{g}\,,\,\,\,
r_{H}=r_{g}[1+(1-a_{*}^{2})^{1/2}]\,,\,\,\,x_{ms}^2=r_{ms}/r_{g}\,,\,
\nonumber\\
&& x^2=r/r_{g}\,,\,\,\,\xi= x^2/x_{ms}^2\,,\,\,\,
q=(1-a_{*}^{2})^{1/2}\,,\,\,
\beta\equiv\frac{\Omega_{D}}{\Omega_{H}}=\frac{2(1+q)}{a_{*}}[x^{3}+a_{*}]^{-1}\,,
 \label{eq:3414}
\end{eqnarray}
where \emph{r$_{H}$} is the radius of BHs on the horizon. The
mapping of the polar angle with the radius $\xi$ is defined as
\begin{eqnarray}
  \cos\theta &=&\int_{1}^{\xi}G(a_{*},n,\xi)d\xi\,,\,\,\,
\cos\theta_{1}=\int_{1}^{\xi_{out}}G(a_{*},n,\xi)d\xi\,,
\label{eq:3414}
\end{eqnarray}
where $\theta$$_{1}$=$\pi$/6, and
\begin{equation}
  G(a_{*},n,\xi)=-\frac{\xi^{1-n}x_{ms}^{2}\sqrt{1+a_{*}^{2}x_{ms}\xi^{-2}+2a_{*}^{2}x_{ms}^{-6}\xi^{-3}}}
  {2\sqrt{(1+a_{*}^{2}x_{ms}^{-4}+2a_{*}^{2}x_{ms}^{-6})(1-2x_{ms}^{-2}\xi^{-1}+a_{*}^{2}x_{ms}^{-4}\xi^{-2})}}\,.
\label{eq:3414}
\end{equation}
 Since the magnetic field on the horizon of a BH is brought and held by its
surrounding magnetized disk, for a cylindrical magnetic field with
both poloidal and toroidal components, the poloidal magnetic field
on the disk is assumed to vary with $\xi$ as a power law
(Blandford 1976; Wang et al.~\cite{wang02},~\cite{wang03})
\begin{eqnarray}
B_{disk}&\propto& \xi^{-n}\,\,,
\end{eqnarray}
where $n$ is the power law index. The distribution of the angular
momentum flux $H$ transferred between the BH and the disk is (Wang
et al.~\cite{wang02},~\cite{wang03})
\begin{equation}
  H= \left\lbrace \begin{array}{ll}~H_{0}A(a_{*},\xi)\xi^{-n} ~~~~~~1<\xi<\xi_{out}\\
                                              ~0 ~~~~~~~~~~~~~~~~~~~~~~~~\xi>\xi_{out}
                             \end{array} \right.
\label{eq:Harr}
\end{equation}
where
\emph{H}$_{0}$=$<$\emph{B}$_{BH}$$>^{2}$M=1.48$\times$10$^{21}$\emph{B}$_{4}$$^{2}$\emph{M}$_{8}$\emph{
g s$^{-2}$}; \emph{B}$_{4}$=\emph{B/\rm{10}$^{4}$ \rm{G}},
\emph{M$_{8}$}=\emph{M/\rm{10}$^{8}$M$_{\odot}$}, and $B_{BH}$ is the magnetic field on the BH horizon.\\

Substituting \emph{F}$_{DA}$ and \emph{F}$_{MC}$ into Eq.(1), we
can obtain \emph{T}(\emph{x}). Then we can estimate the energy
flux at frequency $\nu$ from the disk, which is
\begin{equation}
 \nu L_{\nu}=\nu \int_{r_{in}}^{r_{out}}2\pi I_{\nu}rdr\,,
\label{eq:3414}
\end{equation}
where $I_{\nu}$ is the flux emitted by each unit disk element,
which is assumed, very crudely, to satisfy
\begin{equation}
 I_{\nu}=B_{\nu}[T(x)]=\frac{2h\nu^{3}}{c^{2}(e^{h\nu/kT(x)}-1)}
\label{eq:3414}
\end{equation}

Setting the boundary condition and assuming that the disk extends
down to the last stable orbit of the BH, it is thus possible to
determine the relation of
$\nu$\emph{L$_{\nu}$}(5100\AA)-\emph{M$_{BH}$} from Eq.(1) to
Eq.(12). The $\nu$\emph{L$_{\nu}$}(5100\AA) versus \emph{M$_{BH}$}
relationships as functions of \emph{a$_{*}$},\emph{n} and $B_{BH}$
are plotted in Figure 1, Figure 2 and Figure 3, respectively. The
figures show that for the Kerr black hole the
$\nu$\emph{L$_{\nu}$}(5100\AA)-\emph{M$_{BH}$} relation is not
very sensitive to \emph{a$_{*}$} and \emph{n}, but sensitive to
the magnetic field strength $B_{BH}$. Therefore the magnetic field
strength is the only sensitive parameter for
$\nu$\emph{L$_{\nu}$}(5100\AA)-\emph{M$_{BH}$} relation.\\

\section{The correlation between $B_{BH}$ and \emph{M$_{BH}$}}

Comparing Eq.(11) with the observational relation of $\nu
L_\nu(5100\AA)-M_{BH}$, we can therefore estimate the magnetic
field strength $B_{BH}$ for each source. The data used to get the
$\nu L^{obs}_\nu(5100\AA)$ relation here include 29 objects of
Kaspi et al.'s\,(\cite{kasp00}) samples, 30 objects of McLure et
al.'s(~\cite{mclu01}) samples, and 84 quasars of Shields et
al.'s(~\cite{shie03}) samples. We omitted 5 objects of Kaspi et
al.'s (\cite{kasp00}) since their lower uncertainties are too
large. A correlation between the black hole masses and the
magnetic field strength is obtained for the 143 AGN sources
discussed above, which is
\begin{eqnarray}
\log {B_{BH}}=9.26-0.81\log{M/M_\odot}\,\,\,.
\end{eqnarray}
The derived \emph{B} versus \emph{M$_{BH}$} correlation is plotted
as dashed-line in Figure 4, together with data used.

This relationship between ${M}_{BH}$ and $B_{BH}$ is significantly
differ with the relationship based on the standard accretion disc
theory (Shakura \& Sunyaev ~\cite{ss73}; Novikov \& Throne
~\cite{nt73}) which is $B\sim10^{8}{M/M_{\odot}}^{-1/2}G$ ground
on the balance between the radiation pressure and magnetic
viscosity pressure for the case of high accretion rate that are
respectively:
\begin{eqnarray}
&&
P_{r}=2\times10^{16}(\alpha\frac{M}{M_{\odot}})^{-1}(2r_{ms})^{-\frac{3}{2}}
\,,\,\,\, P_{mag}=B^{2}/8\pi\,\,
\end{eqnarray}

\section{Summary}
\label{sect:discussion} Based on the MC process which can transfer
energy and angular momentum from a fast-rotating Kerr BH to its
surrounding disk and then radiated in the way of thermal emission,
we investigate the relation of the BH mass with its magnetic field
by comparing with observation data, provided that the magnetic
field on the black hole horizon is constant and the poloidal
magnetic field on the disk varies with the radius as power law.
The model shows that the relation of
$\nu$\emph{L$_{\nu}$}(5100\AA)-\emph{M$_{BH}$} is not very
sensitive on the values of \emph{a$_{*}$} and \emph{n}. The value
of $B_{BH}$ on the horizon of Kerr BH is the only sensitive
parameter for the emission of $\nu L_\nu(5100\AA)$ from the disk.
Consequently a correlation $\log {B}=9.26-0.81\log{M/M_\odot}$ is
obtained for the samples of 143 AGN sources. This relation is
different with the result ground on the standard accretion disc
situation at high accretion rate. The model is limited by the
available data $\nu L_\nu(5100\AA)$ from the samples of Kapspi et
al.(~\cite{kasp00}), McLure et al.(~\cite{mclu01}), and Shields et
al.(~\cite{shie03}).

\label{lastpage}

\end{document}